\documentclass[twocolumn,prl,nofootinbib,showpacs]{revtex4-1}
\usepackage{graphicx}
\usepackage{hyperref}
\usepackage{amsfonts}
\usepackage{amsmath,amssymb}
\usepackage{bm}

\begin{document}

\title{Superconducting proximity effect in semiconductor nanowires}
\author{Tudor D. Stanescu}
\affiliation{Department of Physics, West Virginia University, Morgantown, WV 26506, USA}
\author{S. Das Sarma}
\affiliation{Condensed Matter Theory Center, Department of Physics, University of Maryland, College Park, MD 20742, USA}

\begin{abstract}
We theoretically consider the proximity effect in semiconductor--superconductor hybrid nanostructures, which are being extensively studied in the context of the ongoing search for non--Abelian Majorana fermions in solid state systems. Specifically, we consider the dependence on the thickness of the semiconductor in the direction normal to the interface,  a physical effect that has been uncritically neglected in all prior work on the subject. Quite surprisingly, we find the completely unanticipated result that increasing the semiconductor thickness leads to a drastic suppression of the induced superconducting gap due to proximity--induced interband coupling. As a result, in the limit of strong semiconductor--superconductor coupling, the proximity--induced gap becomes much smaller than the bulk superconductor gap and depends weakly on the interface transparency. 
\end{abstract}


\maketitle

{\em Introduction}. The concrete prediction~\cite{Sau2010,Alicea2010,Sau2010a,Lutchyn2010,Oreg2010} about the existence of zero--energy localized Majorana modes  in proximity--coupled semiconductor--superconductor (SM--SC) hybrid structures has triggered a great deal of research activity aimed at realizing the Majorana modes and understanding the details of the underlying physics~\cite{Beenakker2011,Alicea2012,Leijnse2012,Stanescu2013}. Very recent experimental reports~\cite{Mourik2012,Das2012,Deng2012,Rokhinson2012,Finck2012} providing observational evidence for the possible realization of the predicted Majorana zero--mode in semiconductor nanowires with proximity--induced superconductivity~\cite{Sau2010a,Lutchyn2010,Oreg2010} have only accelerated the research activity in this field, leading to hundreds of papers in the last 2-3 years. In particular, the nature of the experimentally observed zero--bias peak in the differential tunneling conductance~\cite{Mourik2012,Das2012,Deng2012}, a theoretically--predicted~\cite{Sau2010a,Sengupta2001,Law2009,Flensberg2010} signature of the elusive Majorana mode, is currently actively debated in the literature~\cite{Potter2010,Akhmerov2011,Potter2011,Lutchyn2012,Lin2012,Stanescu2012,DSarma2012,Prada2012,Bagrets2012,Liu2012,Pikulin2012,Kells2012}. 

The theoretical work presented in this paper takes a step back from the hot issue concerning the nature of the observed zero--bias conductance peak (ZBCP) in the topological superconducting phase and addresses a more basic question: What are the characteristic features of the superconducting proximity effect in SM--SC nanostructures and how does proximity--induced superconductivity depend on the geometrical details of the SM component, in particular on the SM thickness in the direction perpendicular to the interface? We believe that a deeper understanding of the nature of proximity--induced superconductivity itself is essential to making further progress in this field. It is surprising that, although many effects that depend on details of the SM--SC structure (e.g., various types of disorder, multiband occupancy, length of the SM wire and overlap between the Majorana modes localized at opposite ends, soft gap, thermal broadening of the Majorana mode, decay of the zero--mode due to coupling to the tunneling contacts, possible braiding architectures, etc.)  have been discussed in the literature in some depth, the invariably present finite thickness effect corresponding to the finite dimension of the SM layer in the direction transverse to the SM--SC interface has never been accounted for. In this work, we develop a theory  for the proximity effect in SM--SC hybrid nanostructures that explicitly takes into account the multiband nature of the SM spectrum due to the finite thickness of the SM layer, $L_z=N_z a$, where $a$ is the lattice constant and $N_z$ the number of atomic monolayers in the direction normal to the interface, and incorporates the effects of proximity--induced inter--band coupling. We emphasize that earlier theories have considered the unphysical limit $L_z\rightarrow 0$, when the SM has no band structure arising from the $z$--confinement, while,  in reality, the experimental SM--SC hybrid structures used so far in the laboratory have $L_z\approx 50-100$nm.

We find the qualitatively important and unexpected result that, for finite $L_z$ above a critical value $L_0$ that depends on the parameters of the materials (typically $L_0\sim 40$nm for InSb-- and InAs--based structures), the proximity--induced SC gap becomes strongly suppressed, depends non--monotonically on the effective SM--SC tunnel coupling $\gamma$, and vanishes in the strong coupling limit. This is in complete contrast with the thin layer limit $L_z\rightarrow 0$, the only case studied so far in the literature, where the proximity--induced gap increases monotonically with the effective SM--SC coupling up to the maximum value given by the bulk gap of the SC system.  The mechanism responsible for this surprising result is generated by the proximity--induced coupling among the SM bands produced by the quantum mechanical confinement in the direction transverse to the interface. When $L_z$ is large enough, so that the confinement--induced inter--band gaps become comparable to the effective SM--SC coupling, the inter--band coupling has a very strong, nonperturbative effect on the induced SC gap, and, as a result, the induced gap is suppressed. 

The finite transverse thickness--induced suppression of the SC gap may provide an explanation for why the proximity--induced gaps corresponding to specific SM/SC materials measured in different samples and different laboratories have almost identical values. This is a rather puzzling situation  from the perspective of a $L_z\rightarrow 0$ theory:  The induced gap is much smaller than the bulk SC gap $\Delta_0$, hence weak effective coupling across the SM--SC interface is required, $\gamma \ll \Delta_0$. However, in this limit the proximity gap is proportional to $\gamma$, which is expected to show substantial sample to sample variation. Our current work shows that in SM systems with $L_Z>L_0$ a proximity gap much smaller that $\Delta_0$ can occur in the strong coupling limit, where the dependence of the induced gap on $\gamma$ is weak, leading to an apparent universality of the proximity effect in samples of similar thickness.  

{\em Model and theory}. We use a minimal model consisting of an infinite SM wire along $x$-direction with rectangular cross section $L_y\times L_z$ proximity coupled across a $z=0$ interface to an s--wave superconductor that fills the $z<0$ half--space. Our goal is to explicitly include the size quantization effect arising from finite $L_z$ into the theory of the SC proximity effect. Since our main result is independent on $L_y$, we focus on the quasi--2D limit, $L_y\rightarrow \infty$.  In the presence of proximity--induced superconductivity, the low--energy physics of the SM nanowire is described by a Green function matrix with an inverse given by~\cite{Stanescu2011,Stanescu2013}
\begin{equation}
[G^{-1}]_{\bm n \bm n^\prime}(\omega) = \omega - H_{\bm n \bm n^\prime} - \Sigma_{\bm n \bm n^\prime}(\omega),  \label{G1}
\end{equation}
where $H_{\bm n \bm n^\prime}$, with ${\bm n}=(n_y, n_z)$, is the effective low--energy Hamiltonian for the SM nanowire in the basis of the transverse wave functions $\phi_{n_w}(w)=\sqrt{2a/L_w}\sin[w n_w\pi/L_w]$, with $w\in\{y, z\}$.  In Eq. (\ref{G1}) $\Sigma_{\bm n \bm n^\prime}(\omega)$ is the proximity--induced self--energy given by~\cite{Stanescu2011,Stanescu2013}
\begin{equation}
\Sigma_{\bm n \bm n^\prime} = -\gamma_{\bm n \bm n^\prime}\left[\frac{\omega + \Delta_0 \sigma_y\tau_y}{\sqrt{ \Delta_0^2-\omega^2}} + \zeta \tau_z\right], \label{Sigmannp1} 
\end{equation}
where  $\sigma_i$ and $\tau_i$ are Pauli matrices in the spin and Nambu spaces, respectively, and $\zeta$ is a constant that depends on the details of the SC band structure. The effective SM--SC coupling at the interface is represented by the matrix 
$\gamma_{\bm n \bm n^\prime} = g \phi_{n_z}(0)\phi_{n_z^\prime}(0)\delta_{n_y n_y^\prime}$,
where $g=\nu_F|\tilde{t}|^2$ is a constant that depends on the surface density of states of the SC in the normal state ($\nu_F$) and on the transparency of the SM--SC interface ($\tilde{t}$). We have assumed that the coupling at the interface is uniform, i.e., independent of position in the $y$--direction, which results in a block--diagonal structure of the self--energy with respect to $(n_y, n_y^\prime)$. Note that all confinement--induced modes $n_z$ perpendicular to the interface are coupled. The eigenvalues of the low--energy states can be obtained by solving the Bogoliubov--de Gennes (BdG) equation 
\begin{equation}
{\rm det}[G^{-1}(\omega)] = 0, \label{BdG1}
\end{equation}
where the frequency is restricted to values inside the bulk SC gap, $|\omega|<\Delta_0$. To the best of our knowledge, all existing theoretical and numerical work for the SM--SC hybrid nanostructures is based on solutions of Eq. (\ref{BdG1}) that neglect the inter--band coupling  $\gamma_{n_z n_z^\prime}$. While this decoupled band approximation is  expected to be valid in the strong $z$--confinement limit, when the confinement--induced inter--band gaps are much larger than the proximity--induced inter--band couplings, it breaks down for large-enough $L_z$. Note that in the decoupled band approximation  the matrices describing the proximity effect become band--diagonal, e.g., $\gamma_{\bm n \bm n^\prime}=\gamma_{\bm n}\delta_{\bm n \bm n^\prime}$. Furthermore, in the static limit corresponding to $\sqrt{\Delta_0^2 - \omega^2}\approx|\Delta_0|$ in Eq. (\ref{Sigmannp1}), the proximity--induced SC pairing potential becomes $\Delta_{\bm n} = \gamma_{\bm n}\Delta_0/(\gamma_{\bm n}+\Delta_0)$. This limit has been extensively used for discussing the proximity effect in SM--SC hybrid systems. We note that, in this approximation, $\Delta_{\bm n}\approx\Delta_0$ for $\gamma_{\bm n}\gg\Delta_0$ (strong coupling) and $\Delta_{\bm n}\approx\gamma_{\bm n}$ for $\gamma_{\bm n}\ll\Delta_0$ (weak coupling)~\cite{Stanescu2011,Sau2010b}. In the present work, we relax the strong $z$-confinement approximation and consider the interband--coupling effect arising from the $n_z$ bands corresponding to a finite, realistic thickness $L_z$.  

The main consequence of relaxing the strong confinement approximation is that bands with arbitrary $n_z$ values become coupled, which generates second order processes involving low--energy initial and final $n_z$ states and  high--energy  intermediate $n_z$ states. The transitions between these states are proximity--induced and involve hopping into the SC. These processes strongly renormalize the bare SM band parameters and the induced SC pairing potential. In particular,  for a given diagonal coupling strength $\gamma_{\bm n \bm n}$, the inter--band coupling results in the collapse of the induced SC gap in thick--enough nanostructures.  Also, we note that, in this regime, the construction of an effective Hamiltonian that describes the SM--SC nanostructure has to involve the high--energy $n_z$ bands, as they are intrinsically coupled to the low--energy bands and renormalize them strongly, which ultimately leads to the collapse of the induced SC gap. Thus, assuming that the higher--lying $n_z$ bands do not matter, as has been universally done in the literature, is simply incorrect and leads to qualitatively wrong results.

\begin{figure}[tbp]
\begin{center}
\includegraphics[width=0.48\textwidth]{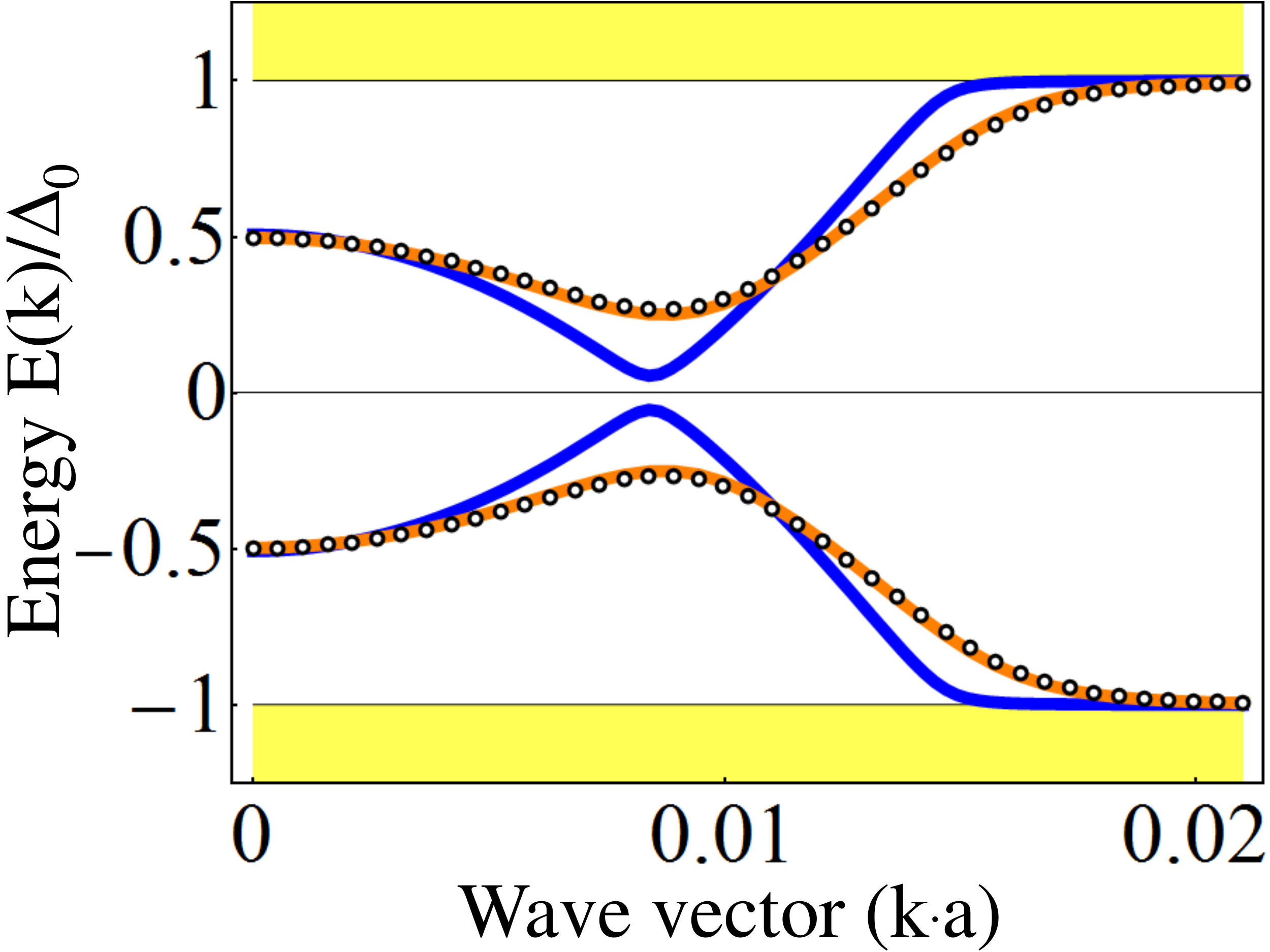}
\vspace{-4mm}
\end{center}
\caption{(Color online) BdG spectra in the vicinity of $k=0$ for 2D semiconductor--superconductor heterostructures with effective coupling $\gamma\approx0.33\Delta_0$ and different values of the semiconductor thickness: $L_z=20$nm (black circles), $L_z=40$nm (orange), and $L_z=80$nm (blue). Two semiconductor bands are partially occupied ($n_z=2$). The yellow regions correspond to bulk superconductor states. In the decoupled--band approximation, the induced gap $\Delta(\gamma, \Delta_0)$ is independent of the semiconductor thickness, but for large $L_z$, proximity--induced inter--band coupling determines the collapse of the induced gap (blue line).}
\vspace{-4mm}
\label{Fig1}
\end{figure}

{\em Results}. To understand the effect of proximity--induced coupling involving different transverse confinement bands, we focus on the matrix structure of the BdG equation corresponding to the $n_z$ quantum numbers  and consider a SM thin film -- SC heterostructure with uniform coupling across the planar interface. Note that the results for the proximity--induced gap will be exactly the same for a nanowire with finite width $L_y$, provided the SM--SC coupling is uniform. Non--uniform coupling effects were previously discussed in the literature~\cite{Stanescu2011} in the strong $z$-confinement approximation. The parameters used in the numerical calculations roughly correspond to the NbTiN--InSb hybrid system measured in the original 
Delft experiment: effective SM mass $m=0.016m_e$, lattice constant $a=0.65$nm,  and bulk pairing potential $\Delta_0=1.5$meV.  To limit the number of parameters, we consider the case of vanishing spin--orbit coupling and  zero Zeeman splitting, but we checked numerically that the results are generic. 
The numerical results presented in this paper are based on a two--band tight--binding model of the SM~\cite{Stanescu2013}, but we have also done similar calculations using a more elaborate 8--band Kane--type model~\cite{Stanescu2013}.  The conclusions presented here are not modified qualitatively, although  the details of the two models differ significantly, e.g., the 8-band model, which has many more parameters, is characterized by an effective mass that depends on $L_z$, and requires different tunneling parameters for the conduction and the valence bands. We also emphasize that the nontrivial dependence of the proximity gap on the semiconductor thickness in the direction transverse to the interface, the main finding  of our work, represents a generic feature of the superconducting proximity effect that remains valid in the presence of spin--orbit coupling and applied magnetic field, e.g., in the Majorana--carrying SC topological phase of the SM. 

\begin{figure}[tbp]
\begin{center}
\includegraphics[width=0.48\textwidth]{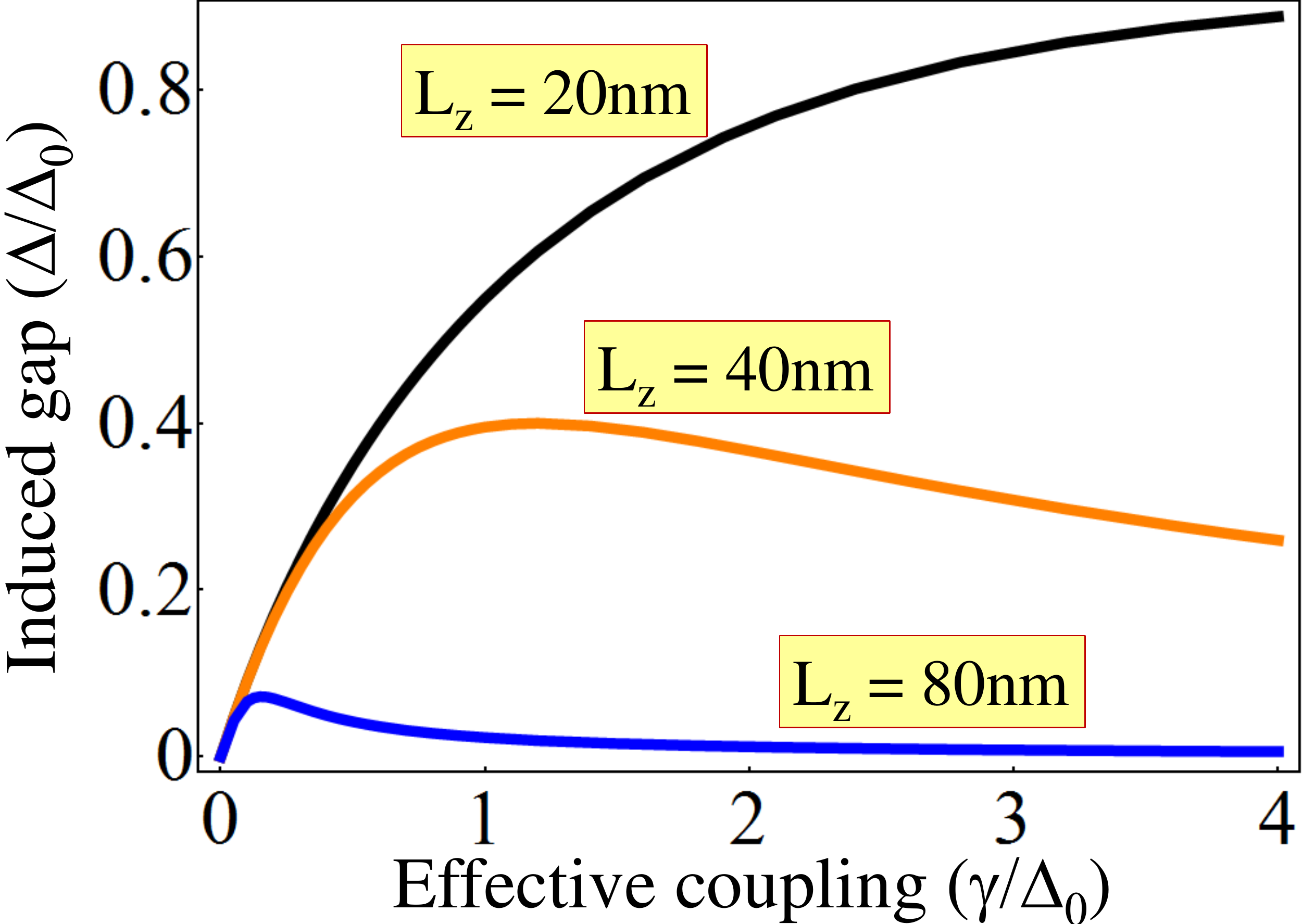}
\vspace{-4mm}
\end{center}
\caption{(Color online) Dependence of the minimum induced gap on the strength of the effective semiconductor--superconductor coupling. In the weak--coupling limit, $\gamma\rightarrow 0$, the minimum induced gap is independent of $L_z$ and is given by the induced pair potential $\Delta_{\rm ind} = \gamma\Delta_0/(\gamma+\Delta_0)$. For large couplings, the minimum induced gap depends on the semiconductor thickness. In thin semiconductors, $\Delta$ increases monotonically and reaches the bulk superconductor value $\Delta_0$ in the strong coupling limit, $\gamma\rightarrow\infty$. By contrast, in thick semiconductors the minimum induced gap depends non--monotonically on the effective coupling and vanishes  in the strong  coupling limit.}
\vspace{-4mm}
\label{Fig2}
\end{figure}

\begin{figure}[tbp]
\begin{center}
\includegraphics[width=0.48\textwidth]{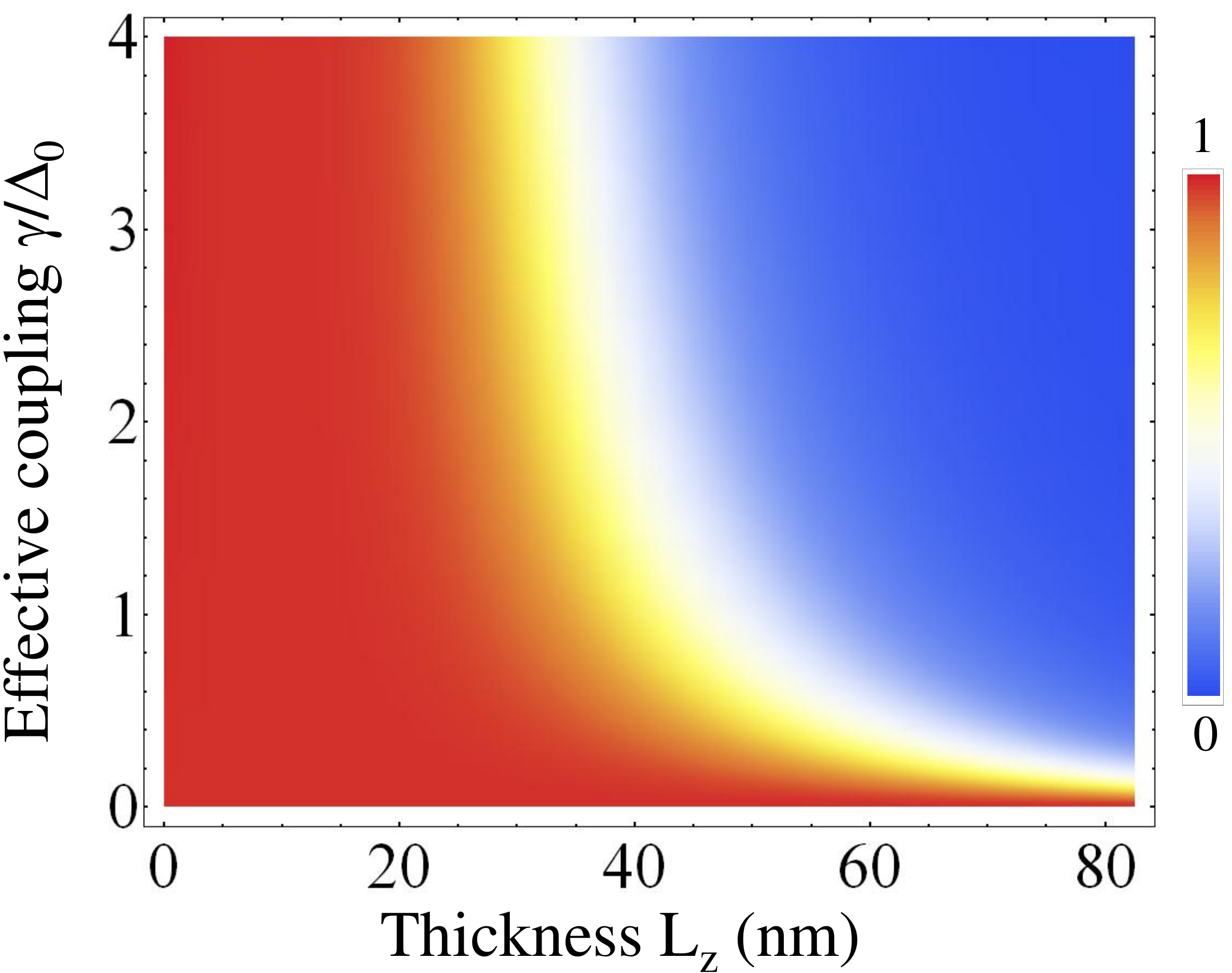}
\vspace{-4mm}
\end{center}
\caption{(Color online) Normalized minimum induced gap, $\delta=\Delta(\gamma, L_z) / \Delta(\gamma, 0)$, as function of the semiconductor thickness $L_z$ and the effective coupling $\gamma$. Thin semiconductors ($L_z<20$nm) are in the decoupled--band regime characterized by a normalized induced gap $\delta\approx 1$ practically independent on $\gamma$. By contrast, the normalized induced gap in thick semiconductors varies from $\delta=1$ in the weak--coupling limit ($\gamma\rightarrow 0$) to $\delta\approx 0$ at strong coupling.}
\vspace{-4mm}
\label{Fig3}
\end{figure}

\begin{figure}[tbp]
\begin{center}
\includegraphics[width=0.48\textwidth]{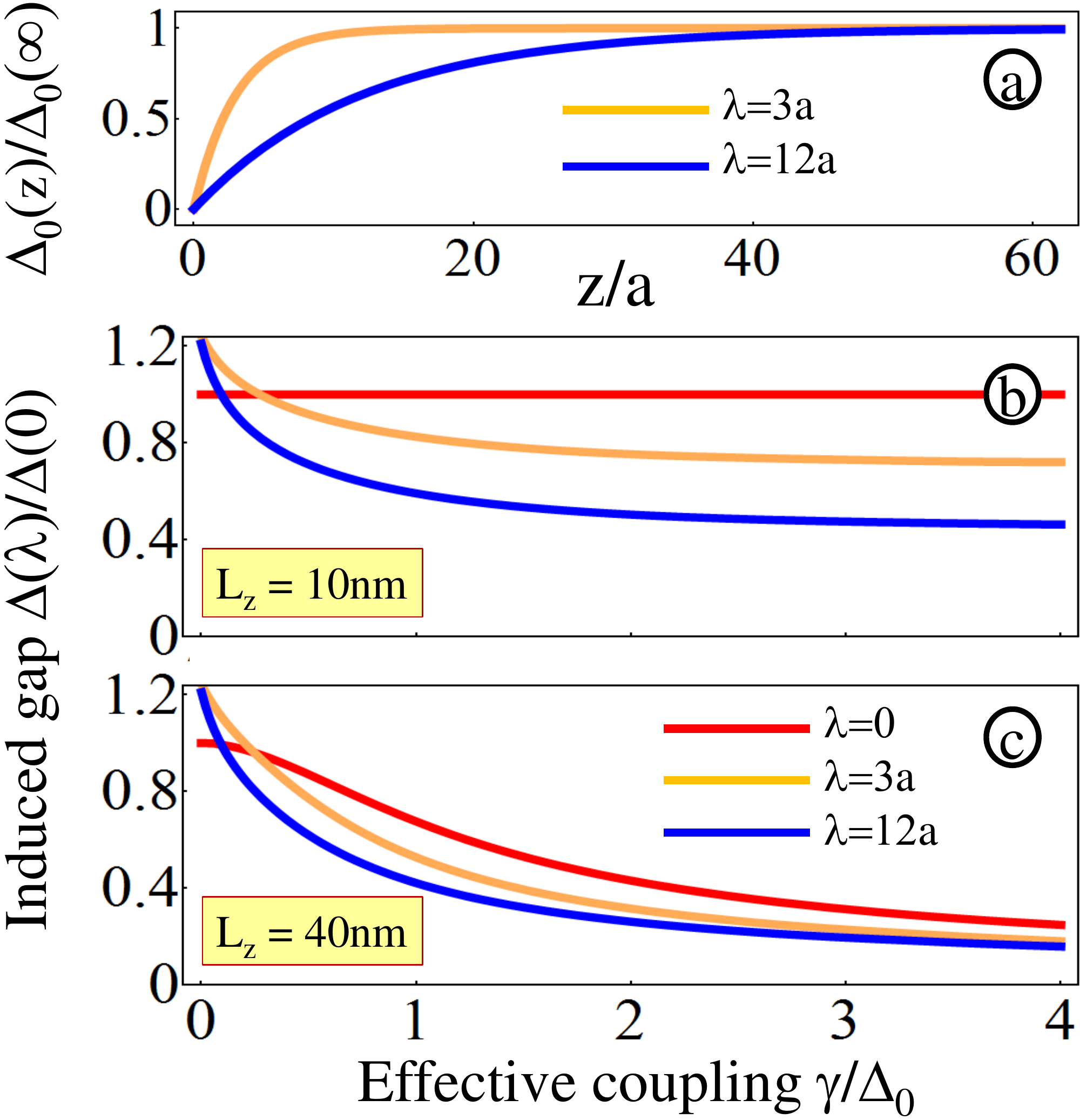}
\vspace{-4mm}
\end{center}
\caption{(Color online) Proximity effect from a bulk superconductor with position--dependent pairing potential. (a) Dependence of the bulk superconductor pairing potential on the distance from the interface. $\Delta_0$ is suppressed on a length scale $\lambda$ and vanishes at the interface. (b) Normalized minimum induced gap, $\delta=\Delta(\gamma, L_z, \lambda) / \Delta(\gamma, 0, 0)$, for $L_z=10$ nm and different values of the characteristic suppression length $\lambda$. (c) Same as in panel (b) for a system with $L_z=40$nm. The red curves ($\lambda=0$) correspond to vertical cuts in the diagram from Fig. 4. Note that, in the weak coupling limit, suppressing $\Delta_0$ near the interface results in a slight increase of the induced gap. For large couplings, the effect of suppressing  $\Delta_0$ near the interface is stronger in thin semiconductors.}
\vspace{-4mm}
\label{Fig4}
\end{figure}

We present our results corresponding to various values of the SM thickness $L_z$ and different SM--SC coupling strengths in Figs. \ref{Fig1}--\ref{Fig4}.  Figure \ref{Fig1} already illustrates clearly our key new finding: The proximity--induced SC gap depends on the thickness of the SM layer and shows a strong suppression for $L_z=80$nm. We emphasize that, in the decoupled band approximation, the induced SC gap corresponding to a given value of the effective coupling $\gamma_{n_z n_z}=\gamma$ is independent of the system size. This approximation holds in the case of strong confinement, as illustrated by the BdG spectra corresponding to $L_z=20$nm and $L_z=40$nm, which are practically identical, but fails for thicker SM layers.
The calculated minimum gap as a function of the effective tunnel coupling across the SM--SC interface for $L_z=20, 40, 80$nm is shown in Fig. \ref{Fig2}. The widely used proximity gap formula $\Delta=\gamma\Delta_0(\gamma+\Delta_0)^{-1}$ represents a good approximation only for very small values of $L_z$,  or for weak effective coupling, but completely fails when the separation of the confinement--induced energy levels $n_z$ becomes comparable with the proximity--induced interband coupling $\gamma_{n_z n_z^\prime}$.  Note that, for $L_z\geq 40$nm and strong--enough coupling, the induced gap is much smaller that the bulk SC gap $\Delta_0$ and depends weakly on $\gamma$. Figure \ref{Fig3} explicitly demonstrates our key finding that, for large $L_z$ ($\geq 40$nm), the induced gap  is strongly suppressed for intermediate and large coupling values and vanishes in the limit $\gamma\rightarrow \infty$, while being described by the ``canonical'' small $L_z$ theory only in the weak coupling limit, $\gamma\rightarrow 0$.
The canonical strong--coupling limit, $\Delta=\Delta_0$, can only be realized in thin SM layers.  In addition, as shown in Fig. \ref{Fig2}, for strong SM--SC coupling and $L_z\geq 40$nm, the induced gap is small and depends weakly on $\gamma$. We believe that this finding explains why the proximity gaps measured in different experimental samples tend to have similar values. Fig. \ref{Fig3} can be viewed as a proximity 'phase diagram' showing the strong suppression of the proximity gap with respect of canonical thin limit value in SM nanostructures with $L_z\geq 40$nm. We note that the crossover value $L_0\approx 40$nm is controlled by the effective mass, which determines the energy spacing between the confinement--induced bands. 

Finally, in Fig. \ref{Fig4} we show the proximity effect in a hybrid system with a SC characterized by a position--dependent pairing potential $\Delta_0(z)$. We assume that the bulk SC pairing potential dies off toward the interface, as expected in a realistic physical situation. Specifically,  $\Delta_0(z)=\Delta_0(\infty)(1-e^{-|z|/\lambda a})$, where $\lambda$ is the characteristic length scale (in units of the lattice constant $a$) for the suppression of the pair potential. The commonly used, and rather unrealistic, theoretical assumption is that the SC pair potential remains constant right up to the interface, i.e., $\lambda=0$. The position dependence of the bulk pairing potential for two values of the healing length $\lambda$ is shown in Fig. \ref{Fig4}(a). The dependence of the normalized induced gap $\Delta(\gamma, L_z, \lambda) / \Delta(\gamma, 0, 0)$ on the effective SM--SC coupling strength for $L_z=10$nm and $L_z=40$nm is shown in  Fig. \ref{Fig4}(b) and (c), respectively. Surprisingly, in the weak--coupling limit, $\gamma\rightarrow 0$, reducing the pair potential near the interface ($\lambda >0$) results in an enhancement of the induced gap with respect to the uniform case $\lambda=0$. For intermediate and strong coupling, increasing the healing length $\lambda$ results in a suppression of the induced gap that is more effective in thin SM layers. 
Adding this finite healing length effect to the finite thickness--induced suppression discussed above results in an expansion of  the 'universal' regime characterized by  small proximity--induced gaps nearly independent of the effective SM--SC coupling toward lower values of the SM thickness. We emphasize that, regardless of $L_z$, this regime corresponds to large tunnel couplings, $\gamma >\Delta_0$. 


We conclude by summarizing our qualitatively new finding: The proximity induced gap in SM--SC hybrid nanostructures is strongly suppressed in the intermediate and strong tunnel coupling regimes whenever the SM layer thickness exceeds a characteristic crossover value determined by the band parameters of the SM. This leads to a strong coupling regime characterized by a small induced gap that is almost tunnel coupling independent, as actually observed experimentally.  If the SC proximity effect in the SM--SC hybrid structures engineered to host the elusive Majorana mode is indeed in the strong coupling regime, a critical reevaluation of the relevant theoretical predictions is urgently needed.

This work is supported by JQI-NSF-PFC, Microsoft Research, DARPA QuEST, and WV HEPC/dsr.12.29.

\bibliography{ProximityRef}

\end{document}